\documentclass[journal,onecolumn]{IEEEtran} 

\usepackage{amsmath,amsthm}
\usepackage{graphics}

\usepackage{dsfont}
\usepackage[dvips]{epsfig}
\usepackage{algorithm, algorithmic}

\usepackage{amssymb}

\newtheorem{lemma}{Lemma}

\newtheorem{defi}{\bf Definition}
\newtheorem{Theorem}{\bf Theorem}

\newtheorem{corol}{\bf Corollary}

\def\Real{\mathop{\hbox{\mit I\kern-.2em R}}\nolimits}
\def\Zeal{\mathop{\hbox{\mit Z\kern-.29em Z}}\nolimits}
\def\be{\begin{equation}}
\def\ee{\end{equation}}

\def\ba{\begin{eqnarray*}}
\def\ea{\end{eqnarray*}}

\def\rank{{\rm rank}}
\def\row{{\rm row}}
\def\col{{\rm col}}
\def\dim{{\rm dim}}
\def\bit{\bibitem}

\input{epsf}
\ifCLASSINFOpdf
\else
\fi
\begin{document}

%

\title{Imperfect Secrecy in Wiretap Channel II} 
\author{Fan Cheng,~\IEEEmembership{Member,~IEEE,}
        Raymond W. Yeung,~\IEEEmembership{Fellow,~IEEE}, and Kenneth W. Shum,~\IEEEmembership{Member,~IEEE}
\thanks{F. Cheng  is with the Institute of Network Coding, The Chinese University of Hong Kong, N.T., Hong Kong. Email: fcheng@inc.cuhk.edu.hk}
\thanks{R. W. Yeung is with the Institute of Network Coding and Department of Information Engineering, The Chinese University of Hong Kong, N.T., Hong Kong.  Email: whyeung@ie.cuhk.edu.hk}
\thanks{ K. W. Shum is with the Institute of Network Coding, The Chinese University of Hong Kong, N.T., Hong Kong.  Email: wkshum@inc.cuhk.edu.hk}
\thanks{This work was partially funded by a grant from the University Grants Committee of the Hong Kong Special Administrative Region (Project No.\ AoE/E-02/08) and Key Laboratory of Network Coding, Shenzhen, China (ZSDY20120619151314964). This paper was presented in part at  ISIT, 2012 \cite{CYSISIT2012}.}
}

\maketitle

\begin{abstract}
In a point-to-point communication system which consists  of a sender, a receiver and  a set of noiseless channels, the sender wishes to transmit a private message to the receiver  through the
channels which may be eavesdropped by a wiretapper. The set of wiretap sets is arbitrary.   The wiretapper
can access any one but not more than one wiretap set.  From each wiretap set, the wiretapper can obtain some
partial information about the private message which is measured by
the equivocation of the message given the symbols obtained
by the wiretapper. The security strategy is to encode the message
with some random key at the sender. Only the message is required to be recovered at the receiver.
Under this setting, we define an achievable
rate tuple consisting of the size of the message, the size of the key, and the equivocation for each wiretap set. We first prove a tight rate region when both the message and the key are required to be recovered at the receiver. Then we extend the result to the general case when only the message is required to be recovered at the receiver.
Moreover, we  show that even if  stochastic encoding is employed at the sender,  the message rate cannot be increased.
\end{abstract}


\section{Introduction}

   \IEEEPARstart{S}{hannon} launched information-theoretic security in his
    seminal paper \cite{shannon1998communication}, where a sender wishes to transmit a private message to a receiver with
    the existence of a wiretapper. The model, referred to as the \textit{Shannon cipher
    system}, requires that the wiretapper can obtain no information
    about the message. In this paper, we  refer to it as \textit{perfect
    security} for ease of discussion. To protect the message,
    the sender encodes the message with a random key which is shared
    with the receiver a priori but unknown to the wiretapper. The sender transmits
    the encrypted message in a public channel to the receiver such that the receiver can recover the message from the key and the encrypted message, while  the wiretapper who  observes the encrypted message only  can obtain no information about the private message.
    The conclusion in \cite{shannon1998communication}, known as the \textit{perfect secrecy theorem}, states that  the
    size of the key can not be less  than the size of the message if perfect
    security is required. Throughout this paper, the size of a random variable is measured by its Shannon entropy. A recent result by Ho \textit{et al.} in \cite{HoTerence2011}
    proved a stronger bound with the additional assumption that the key is independent of  the message:  in the Shannon cipher system, the size of the key
    is lower bounded by the logarithm of the   cardinality of the support  of  the message alphabet.

   \textit{Secret sharing} was studied by
    by Blakley \cite{blakley1979} and Shamir \cite{shamir1979share}, where an even complex model was introduced.
    Ozarow and Wyner \cite{ozarow1985wire} also studied a similar
    model which they called the \textit{wiretap channel~II}.  In their
    model, information is sent to the receiver through a set of noiseless
    point-to-point channels.  It is assumed that the wiretapper can
    access any one but not more than one set of channels, called a
    wiretap set, out of a collection $\cal A$ of all possible wiretap
    sets, where $\cal A$ is specified by the problem under
    consideration. In \cite{ozarow1985wire}, $\mathcal{A}$ consists
    of all the subsets of  the channel set  with size $r$. The strategy
    to protect the private message is the same as that in
    the Shannon cipher system, namely that a key is employed to randomize the message. Specifically, they proved a lower bound on the size of the key which can be attained by a linear code \footnote{The coding scheme in \cite{ozarow1985wire} was called  a group code, which can be represented as a linear code. See \cite{rouayheb2007wiretap} \cite{NgaiYeung09Hamming} for details.}. This
    result is further generalized in Cheng and Yeung
    \cite{ChengYeung2011NetC} for an arbitrary $\cal A$. They proved a lower bound on the size
    of the key and showed that it can   be also achieved by a linear code.

    \textit{Imperfect secrecy}  was independently studied  by Yamamoto \cite{HYamaCodSha} and Yeung
    \cite{yeung2002information} (p. 116).  The communication
    model in \cite{yeung2002information} is the same as the model described in the Shannon cipher
    system, except that the wiretapper may obtain  partial information
    about the message, which is measured by the mutual information between
    the message  and the symbols  obtained by the wiretapper. The
    \textit{imperfect secrecy theorem} states that this mutual information is lower bounded by the
    difference between the size of the message and the size of the key.
    In \cite{HYamaCodSha}, an inequality equivalent to the imperfect secrecy theorem was used in the proof of converse coding theorems for a multiterminal secrecy system.
    When imperfect security is considered in  a wiretap network $\mathcal{G=(V,E)}$,
    where $\cal V$ is the set of nodes and $\cal E$ is the set of
    channels, Cai and Yeung  \cite{cai2002secure} proved two tight bounds, one on the minimum
    size of the key and the other on the maximum size of the message, provided
    that the collection $\cal A$ of all  possible  wiretap sets
    consists of all subsets of $\cal E$  with size $r$ and the information leakage
    about the message for each wiretap set is  at most $i\log q$,
    where $i$ is a fixed integer satisfying $0\leq i\leq r$ and $q$ is
    the size of the alphabet.

    Xu and Chen  \cite{5205251}  studied  how to communicate securely
    over a network in which each channel may be noisy or noiseless. Their
    model is  a single-source single-sink acyclic planar network without
    network coding  and the communication between the source and the sink is subject to
    non-cooperative eavesdropping on each link, namely $\cal A$ consists of
    all the subsets of the channel set  with a single channel. From each wiretap set in $\cal
    A$, the wiretapper can obtain partial information about the
    message, which is measured by  the equivocation of the
    confidential message given the symbols obtained by the wiretapper.
    They defined an achievable rate tuple consisting of the message rate, the key
    rate  and the  equivocation rate for each wiretap set.
    They proved sufficient conditions in terms of the communication rates and the
    network parameters for provably secure communication, along with an
    intuitive and efficient coding scheme. Furthermore,
    the derived achievable rate  region is tight for
    several special cases. In the following, we refer to
    this model as the \textit{non-cooperative imperfect secrecy system}.

    In this work, we  introduce a security model which generalizes the
    model in \cite{ozarow1985wire}. The communication model is the same as
    that in \cite{ozarow1985wire}. The main difference  is
    that in our model $\cal A$ is arbitrary, and from each wiretap set in $\cal A$,
    the wiretapper can obtain some information about the message. On
    the other hand, our model subsumes the noiseless case of the
    model in \cite{5205251}, since the communication in a
    single-source single-sink network without network coding can be
    simplified as a point-to-point system. We also define an achievable rate
    tuple similar to that in \cite{5205251} and a  tight  rate region is
    proved under this setting. 


   The rest  is organized as follows. First,  we present the problem
    formulation and introduce some related results in Section~\ref{sec:im1}. Then we present
    our main result on the rate region in Section~\ref{sec:im2}. Before proving the main result, we  first establish an achievable subregion in Section~\ref{sec:im3} with the additional requirement that the key is also recovered by the receiver.  The main result is proved in Section~\ref{sec:im4}. In Section~\ref{sec:stochastic-encoder}, we show that the message rate cannot be increased by introducing stochastic encoding at the sender.

\section{Problem Formulation and Related Result}\label{sec:im1}
\subsection{Problem Formulation}
\begin{figure}
\begin{center}
\includegraphics[height= 200pt, width = 500pt]{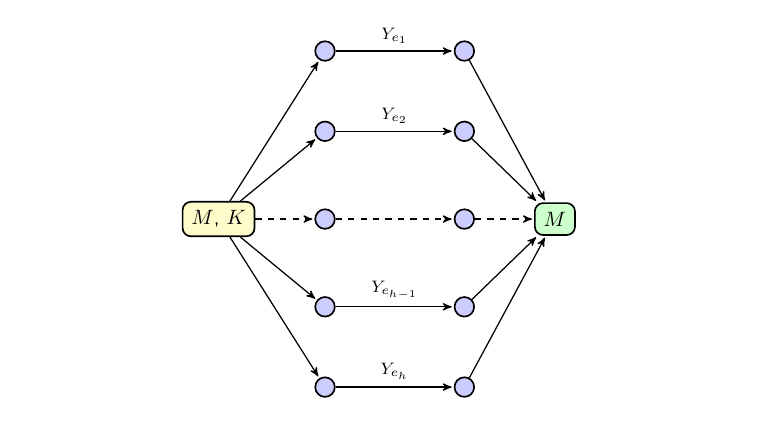}
\caption{Imperfect Wiretap Channel II.}\label{fig-im1}
\end{center}
\end{figure}

The communication model (depicted in Fig. \ref{fig-im1}) in our problem is described as follows:
\begin{itemize}
 \item  The communication is  between a transmitter $s$ and a receiver $t$,
which are connected by a set of point-to-point  noiseless channels.
Let  $\mathcal{E}=\{e_1, e_2, ..., e_h\}$ be the set of channels and
$h= |\mathcal{E}|$. Symbols transmitted on the channels are taken from a common alphabet $\mathcal{F}$ with $|\mathcal{F}|=q$. For each channel $e_i$, $1\leq i\leq h$, the
channel capacity is $C_i\log q$, where $C_i$ is an integer.   Denote the symbols transmitted on
$e_i$ by $Y_{e_i}$.

\item The message $M$ is generated at the transmitter $s$ according
to a uniform distribution on the message set $\mathcal{M}$. The key
$K$, also generated at the transmitter $s$,  takes value in an
alphabet $\mathcal{K}$ according to the uniform distribution, and is
independent of $M$, i.e.,
\begin{equation}
I(M; K) = 0.
\end{equation}
Besides $K$, no additional randomization is allowed inside the network.
The transmitter needs to send  the ciphertext (encrypted
message) to the receiver and the receiver needs to recover  the
message with zero error. Note that the key is only known to the sender. The rates of the message and the key are defined as
follows.
\begin{equation}\label{rate:1}
    R_M= \frac{H(M)}{\log q};
\end{equation}
\begin{equation}
    R_K = \frac{H(K)}{\log q}.
\end{equation}

\item Let $\mathcal{A}$ be the set of  wiretap sets and $|\mathcal{A}|=d$.
 Each wiretapper can access at most one wiretap set in
$\mathcal{A}$. Assume that the wiretapper  knows the encoding and decoding functions but not the private key $K$.

\item For each wiretap set $I_i$, $1\leq i\leq d$, let $Y_{I_i}$ be
the symbols transmitted in $I_i$. It is required that the
wiretapper's equivocation $H(M|Y_{I_i})$ is lower bounded by a given
constant $R_i\log q$, namely
\begin{equation}\label{eqn:equivacation}
 \frac{H(M|Y_{I_i})}{\log q}\geq  R_i.
\end{equation}

\end{itemize}
The achievable rate tuple is defined as follows.
\begin{defi}
The encoder is a  function
$f$ such that
\begin{equation}
f: \mathcal{M}\times \mathcal{K}\to \prod_{i=1}^{h}\mathcal{F}^{C_i}.
\end{equation}
The decoder is a
function $g$ such that
\begin{equation}
g: \prod_{i=1}^{h}\mathcal{F}^{C_i}\to
\mathcal{M}.
\end{equation}
The corresponding rate tuple $(R_M, R_K, R_{i:1\leq i
\leq d})$ is an achievable rate tuple if   $f$ and $g$ satisfy that:
\begin{itemize}
\item [1)] For all $m_1, m_2\in \mathcal{M}$ with  $m_1\neq
m_2$,
\begin{equation}
f(m_1,k_1)\neq f(m_2,k_2),
\end{equation}
for all $k_1, k_2\in \mathcal{ K}$. This guarantees that any two
messages are distinguishable at the receiver; i.e., $$g(f(m,k))=m,$$ for all $k$.
\item [2)] The constraints (\ref{eqn:equivacation}) holds for all $i=1,$ $2,$ $\ldots,$ $d$.
\end{itemize}
\end{defi}

Next, we define the achievable rate tuple  by a block code in terms
of $M$, $K$ and $Y_{I_i},1\leq i\leq d$.

\begin{defi}\label{def:RateTuple}
A rate tuple of  $(R_M, R_K, R_{i:1\leq i\leq d})$ is  achievable by
block codes if there exists a sequence of $(M_n, K_n)$ such that
\begin{align}
&R_M  = \lim_{n \to \infty}\frac{1}{n}\ \frac{\log|\mathcal{M}_n|}{\log q};\label{def1c1}  \\
&R_K   = \lim_{n \to \infty}\frac{1}{n}\ \frac{\log|\mathcal{K}_n|}{\log q}; \label{def1c2}   \\
&R_i  \leq \liminf_{n \to \infty}\frac{1}{n}\
\frac{H(M_n|Y_{I_i,n})}{\log q}, 1\leq i\leq d; \label{def1c3}
\end{align}
where $M_n\in\mathcal{M}_n\subseteq \mathcal{M}^n$,  $K_n\in
\mathcal{K}_n\subseteq \mathcal{K}^n$, and $Y_{I_i,n}\in
\mathcal{F}^n$.
\end{defi}
The inequality  (\ref{def1c3}) means that, for any positive real number $\varepsilon$, there exists a positive integer $n_0$ such that
$$R_i-\varepsilon\leq \frac{1}{n}\frac{H(M_n|Y_{I_i,n})}{\log q}$$
for all $n\geq n_0$.

The rate region $\mathcal{R}$ is defined as the set of all
achievable rate tuples $(R_M, R_K, R_{i:1\leq i\leq d})$. In the
sequel, we refer to this model as the \textit{cooperative imperfect
secrecy system}.

In the sequel, we assume that the base of the logarithm  in the
entropy quantities $(e.g., H(X), I(X;Y))$  is $q$, so that the factor
$(\log q)^{-1}$ can be omitted in (\ref{rate:1})-(\ref{def1c3}).

\subsection{Related Result}

\subsubsection{Perfect  and Imperfect Secrecy}

The perfect secrecy theorem in \cite{shannon1998communication} is
stated as follows.

\begin{Theorem}[Perfect Secrecy Theorem]
    Let $X$ be the plaintext, $Y$ be the ciphertext, and $K$ be the
    key in a secret key cryptosystem. If perfect secrecy is achieved, i.e., $I(X;Y)=0$,
    then
    \begin{equation}\label{perfectSecrecyEqn}
        H(K)\geq H(X).
    \end{equation}
\end{Theorem}

For a network $\mathcal{G}$$=$$(\mathcal{V}, \mathcal{E})$, we denote a cut of $\cal G$
by $(W, W^c)$, where $W\subseteq V$ contains the source node $s$ and $W^c=V\setminus W$
contains the destination node $t$, and refer to  the set of edges from
$W$ to $W^c$  as the cut-set.

For the wiretap network model \cite{cai2002secure}, the following
result related to the perfect secrecy theorem  was proved.
\begin{Theorem}\label{thmK} 
In a wiretap network, let $K$ be the key and $Y_I$ be the symbols
transmitted in  wiretap set $I$. Then
    \begin{equation}\label{impsec:1}
    H(K)\geq H(Y_I).
    \end{equation}
If $I$ is contained in a cut-set $W$, then
    \begin{equation}\label{impsec:2}
    H(M)\leq H(Y_{W\setminus I} |Y_I).
    \end{equation}
\end{Theorem}

As a generalization of the perfect secrecy theorem, the imperfect
secrecy theorem  in \cite{yeung2002information} (p. 116) is stated
below.
\begin{Theorem}[Imperfect Secrecy Theorem]\label{ImperfectSecrecyThm}
    Let $X$ be the plaintext,
    $Y$ be the ciphertext, and $K$ be the key in a secret key
    cryptosystem. Then
    \begin{equation}\label{ImperfectSecrecyEqn}
    I(X;Y)\geq H(X) - H(K).
    \end{equation}
\end{Theorem}
In the above theorem, if  $I(X;Y)=0$, then
(\ref{ImperfectSecrecyEqn})  becomes (\ref{perfectSecrecyEqn}),
i.e., the theorem reduces to the perfect secrecy theorem.  In
\cite{HYamaCodSha},  it was proved that for any secret key
cryptosystem,
\begin{equation}
    H(K)\geq H(X|Y),
\end{equation}
which is equivalent to (\ref{ImperfectSecrecyEqn}).
\subsubsection{ Secure Coding over Routing Networks}
The system model in \cite{5205251} is a single-source single-sink
directed acyclic network with the assumption that each wiretapper
can access only one channel and there is no network coding in the
network. Each channel in the network may be noisy or noiseless.

When all the channels in the network are noiseless, the network  can
be simplified as a point-to-point communication system, in which
each channel is a path from the source node to the destination node
in the original network and the set of wiretap sets $\mathcal{A}$ is
arbitrary.  Hence our model subsumes the non-cooperative model for
this special case.

In \cite{5205251}, an achievable rate region of rate tuples  was
obtained for noisy channels, and the region was shown to be tight
for several special cases. Based on the achievable rate region, they
also gave an algorithm for constructing a secure code on the
network.

The achievable rate region  for noiseless channels is stated below.
\begin{Theorem}[Theorem 2, \cite{5205251}]
    A rate tuple $(R_M, R_K, R_{e}),$ $e\in \mathcal{E}$, is
    achievable if $$0\leq R_e \leq R_M$$ for all $e \in \mathcal{E}$ and  there exist auxiliary numbers $r_{e}$ such that
    \begin{align}
    0 & \leq  r_{e} \leq R_M+R_K;\notag \\
    0 & \leq  R_M+R_K \leq \min_{\textit{Cut}}\sum\limits_{e\in \mathcal{E}_{\textit{Cut}}} r_{e};\notag \\
    0  & \leq  r_{e}  \leq  C_{e}; \notag\\
    R_{e} & \leq  R_M+R_K-r_{e}.\notag
    \end{align}
\end{Theorem}
In the above, $R_{e}$ and $C_e$ correspond to $R_i$ and $C_i$ in our
formulation respectively; 
$\mathcal{E}_{\textit{Cut}}$ is the set of  channels across a given cut
$\textit{Cut}$.

\section{The Rate Region }\label{sec:im2}
The main result of this paper is a characterization of the rate
region $\mathcal{R}$ given by  the following theorem.
\begin{Theorem}\label{General:ratetuple}
    A rate tuple $(R_M,R_K, R_{i:1\leq i \leq d})$ is in $\mathcal{R}$ if and only if
    \begin{align}
    R_M  &\geq R_i, \ 1\leq i \leq d;\label{GRRRT:2}
    \end{align}
    and there exist  $r_i$'s such that
    \begin{align}
    R_K   &\geq \sum\limits_{i=1}^{h}r_i - R_M;\label{GRRRT:1}\\
    R_M &\leq \sum\limits_{i=1}^{h}r_i; \label{GRRRT:3}\\
    0  &\leq r_i \leq C_i,&  \text{$1\leq i\leq h$};\label{GRRRT:4}\\
    0 &\leq R_j \leq  \sum\limits_{e_i\in \bar{I}_j}r_i,& 1\leq j\leq d,\label{GRRRT:5}
    \end{align}
where $\bar{I}_j=\mathcal{E}\setminus I_j$.
\end{Theorem}
Our model is a generalization of the wiretap channel II studied in \cite{ozarow1985wire}, because here we consider imperfect secrecy instead of  perfect secrecy. By letting $\mathcal{A}~=~\{A:~A\subseteq~\mathcal{E},\text{ and }~|A|=r\}$, we can recover the result in \cite{ozarow1985wire}.
Before proving Theorem~\ref{General:ratetuple}, we first study a subregion of
$\mathcal{R}$.

\section{A Subregion of the Rate Region}\label{sec:im3}
By requiring both the message and the key  to be recovered at the
receiver,   we can define a subregion $\mathcal{R}'$ of the rate region $\mathcal{R}$. The  definition of $\mathcal{R}'$ is given below.
\begin{defi}\label{def:subtuple}
The encoder is a  function $f$ such that
\begin{equation}f: \mathcal{M}\times\mathcal{K}\to \prod_{i=1}^{h}\mathcal{F}^{C_i}.\end{equation}
The decoder is a function $g$ such that
\begin{equation}g: \prod_{i=1}^{h}\mathcal{F}^{C_i}\to \mathcal{M}\times\mathcal{K}.\end{equation}
The corresponding rate tuple $(R_M, R_K, R_{i:1\leq i \leq d})$ is a $K$-achievable rate tuple if
$g\circ f$ is the identity function and (\ref{eqn:equivacation}) holds for
all $i=1,$ $2,$ $\ldots,$ $d$.
\end{defi}
The rate region $\mathcal{R}'$ is defined as the set of all
$K$-achievable rate tuples $(R_M, R_K, R_{i:1\leq i\leq d})$.  The region $\mathcal{R}'$ is characterized as follows.
\begin{Theorem}\label{ratetuple}
    A rate tuple $(R_M,R_K, R_{i:1\leq i \leq d})$ is in $\mathcal{R}'$ if and only if
    \begin{align}
    R_M &\geq R_i, \ 1\leq i \leq d;\label{RRRT:2}\\
    R_i &\geq 0,\ 1\leq i \leq d;\label{RRRT:3}\\
    R_K &\geq 0\label{RRRT:4};
    \end{align}
    and there exist  $r_i$'s such that
    \begin{align}
    R_M &= \sum\limits_{i=1}^{h}r_i - R_K;&\label{RRRT:1}\\
    0  &\leq r_i \leq C_i,&   \text{$1\leq i\leq h$};\label{RRRT:5}\\
    \sum\limits_{e_i\in I_j}r_i &\leq R_K+R_M-R_j,& 1\leq j\leq d.\label{RRRT:6}
    \end{align}
\end{Theorem}

\subsection{Converse}\label{sec1:converse}
    In this section, we prove that if $(R_M,R_K, R_{i:1\leq i \leq
    d})\in \mathcal{R}'$, then the constraints (\ref{RRRT:2})-(\ref{RRRT:6})
    hold. Since the converse is valid for both   single-shot coding ($n=1$) and block coding ($n\geq 1$), we  prove it only for  single-shot coding for simplicity. The constraints (\ref{RRRT:3}) and (\ref{RRRT:4}) are
    obvious.

    We first prove the constraint (\ref{RRRT:2}). By the constraint (\ref{eqn:equivacation}),
    \begin{equation}\label{prop1}
    R_i  \leq  H(M|Y_{I_i})\leq  H(M)= R_M.
    \end{equation}
Hence the constraints (\ref{RRRT:2})-(\ref{RRRT:4}) hold.

Let us  consider an equivalent condition of the constraint
(\ref{eqn:equivacation}). For all $1\leq i\leq d$, let
\begin{equation}\label{eqn:rrrt1}
c_i=R_M-R_i=H(M) -R_i.
\end{equation}
The constraint (\ref{eqn:equivacation}) is equivalent to
    \begin{equation}\notag
   I(Y_{I_i};M)\leq H(M)-R_i,
    \end{equation}
or
    \begin{equation}\label{eq20111031}
    0\leq  I(Y_{I_i};M) \leq c_i.
    \end{equation}
By  (\ref{prop1}) and
    (\ref{eqn:rrrt1}),
     $$0\leq c_i\leq R_M.$$

    Next, we prove a lemma which generalizes  the inequality (\ref{impsec:1}) in Theorem~\ref{thmK}.
    \begin{lemma}\label{ImperfectSecureNCThm}
        In a cooperative imperfect secrecy system, let  $M$ be the message,
        $K$ be the key and $Y_I$ be the symbols transmitted  in wiretap set $I$. Then
        \begin{equation}
        I(Y_I;M)\geq H(Y_I) - H(K).
        \end{equation}
    \end{lemma}
    \begin{proof}
    Since $I(M;K) = 0$ and $H(Y_I|M,K) =0$,
    \begin{align}
    I(Y_I;M)& = H(Y_I) - H(Y_I|M) \notag \\
            & \geq H(Y_I) - H(Y_I,K|M)\notag  \\
            & = H(Y_I)-H(K|M)-H(Y_I|K,M)\notag\\
            & = H(Y_I) - H(K|M) \notag \\
            & = H(Y_I) - H(K). \notag
    \end{align}
    \end{proof}

    In the  next theorem, we prove the constraints (\ref{RRRT:1}),
    (\ref{RRRT:5}), and (\ref{RRRT:6}).
    \begin{lemma}\label{thmO}
        For any tuple  $(R_M$, $R_K$, $R_{i:1\leq i\leq d})$ $\in \mathcal{R}'$, there exist $r_i$'s such that
    \begin{align}
        R_M &=  \sum\limits_{i=1}^{h}r_i - R_K;& \notag\\
        0  &\leq r_i \leq C_i,&  \text{$1\leq i\leq h$};\notag\\
        \sum\limits_{e_i\in I_j}r_i& \leq R_K+R_M-R_j,& 1\leq j\leq d.\notag
        \end{align}
    \end{lemma}

    \begin{proof}
        By Lemma \ref{ImperfectSecureNCThm} and the inequality (\ref{eq20111031}),
        for each wiretap set $I_i$,
        \begin{equation}\notag
        H(Y_{I_i}) - H(K) \leq I(Y_{I_i};M) \leq c_i,
        \end{equation}
        or
        \begin{equation}
        H(Y_{I_i})\leq  H(K)+c_i= R_K+c_i.
        \end{equation}
        For each channel $e_i$, $1\leq i \leq
        h$,
        \begin{equation}
        H(Y_{e_i})\leq C_i.
        \end{equation}
        Since
        $Y_{(e_i:1\leq i\leq h)}$ is a function of $(M,K)$ and $(M,K)$ can be recovered by $Y_{(e_i:1\leq i\leq h)}$,
        \begin{equation}\label{eqineq:1}
        H(Y_{(e_i:1\leq i\leq h)})= H(M,K)=H(M)+H(K).
        \end{equation}
        Hence,
        \begin{equation}\notag
        H(M) = H(Y_{(e_i:1\leq i\leq h)}) - H(K),
        \end{equation}
        which is equivalent to
        \begin{equation}
        R_M = H(Y_{(e_i:1\leq i\leq h)}) - R_K.
        \end{equation}
        For $1\leq i \leq h$, let
        $$r_i = H(Y_{e_i}|Y_{(e_1, e_2, ..., e_{i-1})}).$$
        Then for all  $I_j$, $1\leq j\leq d$,
    $$r_i\leq H(Y_{e_i}|Y_{(e_l: e_l\in I_j, l< i)}).$$
Furthermore,
    \begin{align}
    R_M &=  H(Y_{(e_i:1\leq i\leq h)}) - R_K  \notag\\
        &= \sum\limits_{i=1}^{h}H(Y_{e_i}|Y_{(e_1, e_2, ..., e_{i-1})}) -R_K\notag\\
        &= \sum\limits_{i=1}^{h} r_i -R_K;\notag\\
    0\leq r_i &\leq H(Y_{e_i}) \leq C_i;\notag
    \end{align}
 \begin{align}
    \sum\limits_{e_i\in I_j}r_i&\leq\sum\limits_{e_i\in I_j} H(Y_{e_i}|Y_{(e_l: e_l\in I_j, l< i)}) \notag \\
        & =H(Y_{I_j})  \notag\\
    & \leq R_K+c_j\notag\\
    & = R_K+R_M-R_j,\  1\leq j\leq d, \notag
    \end{align}
        which completes the proof.
    \end{proof}


\subsection{Achievability}\label{sec:ach}
 In this section, we prove that $(R_M$, $R_K$, $R_{i:1\leq i\leq d})$ $\in$  $\mathcal{R}'$ if
 there exists $(r_1, r_2, ..., r_h)$ such that the constraints (\ref{RRRT:2})-(\ref{RRRT:6})
 are satisfied.

In the  following,  a special code in which the symbols sent on the
channels are mutually independent is studied. 
We design a block code with length $n$ as follows. The sender
generates  $M$ and $K$ at rates $R_M$ and $R_K$, respectively,  and
sends symbols on each channel $e_i$ ($1\leq i\leq h$) at rate $r_i$.
Next, we prove that the tuple $(R_M$,~$R_K$,~$R_{i:1\leq i\leq d})$
can be attained by a linear
 code.

Let the symbols  on  channel $e_i$ ($1\leq i\leq h$)  be $Y_{e_i}$. 
For simplicity, assume that the quantities $c_i$ (Recall the definition in (\ref{eqn:rrrt1})), $C_i$, $R_M$, $R_K$, and $r_i$ are all rational numbers, so that there is a sufficiently large $n$ such that 
\begin{align}
c_i'&= nc_i;\label{flr1} \\
    C_i'&= nC_i;\label{flr2}\\
n_M&= nR_M  =  nH(M);\label{flr3}\\
n_K& = nR_K  = nH(K);\label{flr4} \\
n_i&= nr_i= nH(Y_{e_i}),\ 1\leq i\leq
h\label{flr5}
\end{align}
are all integers.
Thus, by (\ref{RRRT:1}), (\ref{RRRT:5}), and (\ref{RRRT:6}), $n_M$,
$n_K$, and $(n_1, n_2, ..., n_h)$ satisfy
\begin{align}
n_M &=  \sum\limits_{i=1}^{h} n_i - n_K;\label{neqn3}\\
   0\leq n_i & \leq C_i', 1\leq i\leq h;  \label{neqn2}\\
 \sum\limits_{e_j\in I_i} n_j&\leq n_K+c_i', 1\leq i \leq d. \label{neqn1}
\end{align}


For a matrix $A$, we write the number of rows and columns of $A$ as
$\row(A)$ and $\col(A)$, respectively. The following two lemmas are
instrumental in the subsequent proofs.

\begin{lemma}\label{lem6}
  Let $F_q$ be a finite field of size $q$, $A,$ $B$ be given matrices with the same number of rows and $(A,B)$ be  the concatenated matrix of $A$ and $B$.
  Let $Y=AM+BK$, where  $\rank(A,B) = \row(A,B)$. If  $M$ and $K$ are
    uniformly distributed on $F_q^m$ and $F_q^k$, respectively,  and
    $I(M;K) = 0$, then
    \begin{equation}\notag
     I(Y;M) = \rank(A,B) - \rank(B).
    \end{equation}
\end{lemma}
\begin{proof}
\begin{align}
    I(Y;M) &= H(Y) - H(Y|M)  \notag \\
            & =H(Y) - H(AM+BK|M)  \notag\\
            & = H(Y) -H(BK|M)\notag\\
        & = H(Y)-H(BK)\notag\\
        & = \rank(A,B) - \rank(B).\notag
    \end{align}
\end{proof}

\begin{lemma}[Lemma 3, \cite{cai2002secure}]\label{cccc2}
    Let $V_1$, $V_2$, ..., $V_m$ be  vector subspaces in $F^n_q$, and
    $\dim(V_i)$ $ =d_i$ $(1\leq i\leq m)$. If $d\geq 0$ and $d+d_i\leq n$
    $(1\leq i\leq m)$, then for  $q>m$, there exists a vector subspace
    $V$ of $F^n_q$, such that $\dim(V)=d$ and $\dim(V \oplus V_i) =
    \dim(V)+\dim(V_i)$ $(1\leq i\leq m)$.
\end{lemma}
\begin{proof}
Let $\{b_1,b_2, ..., b_d\}$ be a  basis of $V$. For all $1\leq i\leq m$, let  $\{v_{i1},v_{i2},..., v_{id_i}\}$ be a maximally
independent set of vectors in $V_i$. We construct $\{b_1,b_2, ..., b_d\}$ by induction. It suffices to show that for
$1\leq j\leq d$, if $b_1, b_2, ..., b_{j-1}$ have been chosen such
that for all $V_i$, $1\leq i\leq m$,
\begin{equation}\label{20120206:1}
b_1, b_2, ..., b_{j-1},v_{i1},v_{i2},..., v_{id_i}
\end{equation}
are linearly independent, then it is possible to choose $b_j$ such
that for all $1\leq i\leq m$,
\begin{equation}
b_1, b_2, ..., b_{j-1}, b_j, v_{i1},v_{i2},..., v_{id_i}
\end{equation}
are linearly independent. Specifically, $b_j$ is chosen such that it
is independent of the set of vectors in (\ref{20120206:1}) for all
$1\leq i\leq m$; i.e.,
\begin{equation}
    b_j\in F_q^n\setminus\cup_{1\leq i\leq m}\langle b_1, b_2, ..., b_{j-1},v_{i1},v_{i2},..., v_{id_i}\rangle.
\end{equation}
Since the cardinality of a subspace in $F_q^n$ is finite, we need to
show that the set above is nonempty. Toward this end, consider
\begin{align}
&\bigg|\bigcup\limits_{1\leq i\leq m}\langle b_1, b_2, ..., b_{j-1},v_{i1},v_{i2},..., v_{id_i}\rangle\bigg|  \notag\\
&\leq \ \sum\limits_{1\leq i\leq m}\bigg|\langle b_1, b_2, ..., b_{j-1},v_{i1},v_{i2},..., v_{id_i}\rangle\bigg| \notag\\
&=\ \sum\limits_{1\leq i\leq m}q^{d_i+j-1}  \notag\\
&\leq  \ \sum\limits_{1\leq i\leq m}q^{n-1}\ ({\rm for}\ d_i+j\leq d_i+d\leq n)  \notag\\
&= \ mq^{n-1}. \notag
\end{align}
Therefore,
\begin{align}
&\ \bigg|F_q^n\setminus\bigcup\limits_{1\leq i\leq m}\langle b_1, b_2, ..., b_{j-1},v_{i1},v_{i2},..., v_{id_i}\rangle\bigg|\notag \\
&\geq \ q^n -mq^{n-1}\notag \\
&= \ q^{n-1}(q-m)\notag \\
&> \ 0, \notag
\end{align}
since $q> m$. Hence $b_j$ can be chosen for all $1\leq j\leq m$.
\end{proof}

The remaining of this subsection is largely  about the following theorem.

\begin{Theorem}\label{thminner}
    When $q$ $>$ $|\mathcal{A}|$ is a prime power, if the integer tuple $(n_1,$ $n_2,$ $...,$
    $n_h)$ satisfies (\ref{neqn3})-(\ref{neqn1}), then there exists a
    linear  code such that $H(M^n) = n_M$ and $H(K^n) = n_K$.
\end{Theorem}
\begin{proof}
The code can be constructed as follows.
Let the finite field $F_q$ be the common alphabet of $M$,  $K$ and
 all the channels. The
symbols transmitted on channel $e_i$ ($1\leq i \leq h$) is taken
from $F_q^{n_i}$, which means there are $n_i$ symbols from $F_q$
transmitted on $e_i$. Let $x_1,\ x_2,\ ...,\ x_{n_M+n_K}$ be all the symbols to send,
where the first $n_1$ symbols are sent on $e_1$, the next $n_2$
symbols are sent on $e_2$, so on and so forth, and the last $n_h$ symbols are
sent on $e_h$. We construct $x_i$'s according to their positions in
the sequence.

 Generate $n_K$ mutually independent symbols $K=(k_1, k_2, ..., k_{n_K})$ from
$F_q$. Transmit $K$ at the first $n_K$ positions, i.e.,
\begin{equation}\label{allmost1}
x_i = k_i=b_i\cdot K, 1\leq i\leq n_K,
\end{equation}
where \begin{equation}\label{def:bi}
b_i = (\underbrace{0, 0, ...,
0,}_{i-1} 1, 0, ..., 0).
\end{equation}
Then generate $n_M(=\sum_{i=1}^{h} n_i -n_K)$ mutually independent message
symbols $(m_1, m_2, ..., m_{n_M})$ from $F_q$. For the
remaining $n_M$  positions in $e_i,1\leq i \leq h$, transmit the
encrypted message with the encoding
\begin{equation}\label{2011-1031-1905}
       x_i = m_{i-n_K} +b_i K,\ \  n_K+1 \leq i\leq n_K+n_M,
\end{equation}
where $b_i\in F_q^{n_K}$ is a row vector   to be determined in the
following steps.

We need to construct $\{b_i:n_K+1\leq i \leq n_K+n_M\}$ such that:
\begin{itemize}
\item [(a)] Both $M$ and $K$ can be recovered at node $t$.
\item [(b)] The  constraint  (\ref{eq20111031}) (which is equivalent to
(\ref{eqn:equivacation}))  holds for all the wiretap sets.
 \end{itemize}
 From the previous discussion, we can see that receiver $t$ can
recover $K$ from the symbols in the first $n_K$ positions, and by
 (\ref{2011-1031-1905}), $M$ can be also recovered via
\begin{equation}\notag
    m_{i-n_K}=x_i-b_iK,\ n_K+1\leq i \leq n_K+n_M.
\end{equation}
Hence, the  condition (a) is satisfied by any choice of $b_i$'s. Moreover, it can readily be seen that $x_i$, $1\leq i\leq n_K+n_M$ are mutually independent.

In matrix form, (\ref{allmost1}) and (\ref{2011-1031-1905}) can be written as
\begin{equation}\notag
    \left(
    \begin{array}{c}
        x_1  \\
        x_2  \\
        ...  \\
        x_{n_M+n_K}  \\
    \end{array}
    \right)=\left(\begin{array}{c|c}
        A & B   \\
    \end{array}
    \right)\left(\begin{array}{c}
        M   \\
        K   \\
    \end{array}
    \right),
\end{equation}
where
\begin{equation}\label{construction1}
    \left(\begin{array}{c|c}
        A & B   \\
    \end{array}
    \right)=\left(
    \begin{array}{c|c}
        \bf{0}  & I_{n_K\times n_K} \\\hline
            & b_{n_K+1} \\
        I_{n_M\times n_M}  & ... \\
            & b_{n_K+n_M}\\
    \end{array}
    \right).
\end{equation}
In the above, $\bf{0}$ is an $n_K\times n_M$ zero matrix and
$I_{n_K\times n_K}$ is an $n_K\times n_K$ identity matrix. Recall
that the symbols obtained in wiretap set $I_i=\{e_{i_1},e_{i_2},
..., e_{i_{|I_i|}}\}$ are $Y_{I_i}$, $1\leq i\leq d$. Then
\begin{equation}\notag
    Y_{I_i}= \left(
    \begin{array}{c}
        x_{i_1} \\
        x_{i_2}\\
        ... \\
        x_{i_{|I_i|}} \\
    \end{array}
    \right)=(A_{I_i}|B_{I_i}) \left(
    \begin{array}{c}
        M^n\\
        K^n\\
    \end{array}
    \right),
\end{equation}
where  $A_{I_i}$ and $B_{I_i}$ are the corresponding sub-matrices of
$A$ and $B$, respectively.

We now derive a  sufficient condition for (\ref{eq20111031}) to be satisfied. This condition will be used for the construction of $b_i$'s. Since
$x_1, x_2, ..., x_{n_M+n_K}$ are mutually independent,
\begin{equation}
\rank(A_{I_i},B_{I_i})=\row(A_{I_i},B_{I_i})=\sum\limits_{e_j\in I_i} n_j.\label{eqn:last}
\end{equation}
 By Lemma \ref{lem6},
\begin{align}
    I(Y_{I_i}; M)&=\rank(A_{I_i},B_{I_i})-\rank(B_{I_i}) \notag\\
            &=  \sum\limits_{e_j\in I_i} n_j -\rank(B_{I_i}).\notag
\end{align}
The constraint  (\ref{eq20111031}) is equivalent to
\begin{equation}
    I(Y_{I_i};M)\leq nc_i =  c_i'.
\end{equation}
Hence, it is sufficient to construct $B_{I_i}$ such that
\begin{equation}
    \sum\limits_{e_j\in I_i} n_j-\rank(B_{I_i})\leq c_i',\notag
\end{equation}
or
\begin{equation}\label{neweq:1}
    \rank(B_{I_i}) \geq \sum\limits_{e_j\in I_i} n_j-c_i', \text{ for all }1\leq i \leq d.
\end{equation}

For $\sum_{e_j\in I_i} n_j$, by (\ref{neqn1}),
we obtain  that
\begin{equation}\label{neweq:2}
     \sum\limits_{e_j\in I_i} n_j-c_i'\leq n_K=\col(B_{I_i}).
\end{equation}

By (\ref{eqn:last}),
\begin{align}
    \sum\limits_{e_j\in I_i} n_j-c_i'&= \row(A_{I_i},B_{I_i})-c_i'\\
                                     &=\row(B_{I_i})-c_i' \\
                                     &\leq \row(B_{I_i}).\label{neweq:3}
\end{align}

In summary, by (\ref{neweq:2}) and (\ref{neweq:3}), we have
\begin{equation}\label{star}
\sum\limits_{e_j\in I_i} n_j-c_i' \leq \min\{\row(B_{I_i}), \col(B_{I_i})\}.
\end{equation}
In order for (\ref{neweq:1}) to be satisfied, in light of (\ref{star}), it suffices to construct $b_i$'s such that for
all $i$, $1 \leq i \leq d$,
\begin{align}\label{sec57}
    \rank(B_{I_i})&=\min\{\row(B_{I_i}), \col(B_{I_i})\}\notag\\
    &=\min\{\sum\limits_{e_j\in I_i} n_j, n_K\},
\end{align}
i.e., $B_{I_i}$ is full rank.

The row vectors $b_j$, $1 \leq j \leq n_K$, have been defined according to (\ref{def:bi}). In the following, we will construct $b_j$, $n_K+1\leq j\leq n_K+n_M$, iteratively. For each wiretap set $I_i$, $1\leq i\leq d$ and  for each $j$, $1\leq j\leq n_K+n_M$, let
\begin{equation}\notag
    Y_{I_i}^{j}= \left(
    \begin{array}{c}
        x_{i_1} \\
        x_{i_2}\\
        ... \\
        x_{i_{l}} \\
    \end{array}
    \right)=(A_{I_i}^j|B_{I_i}^j) \left(
    \begin{array}{c}
        M\\
        K\\
    \end{array}
    \right),
\end{equation}
where $x_{i_l}$'s are the symbols in  $I_i$ such that $1\leq i_l\leq
j$. Thus, $Y_{I_i}^j$ is a sub-vector of $Y_{I_i}$ up to the $j$\textit{th} row
and  $A_{I_i}^j$ and $B_{I_i}^j$ are the corresponding sub-matrices of $A_{I_i}$ and $B_{I_i}$, respectively. Also, $Y_{I_i}^{j}$, $A_{I_i}^j$ and $B_{I_i}^j$ are  sub-vectors of $Y_{I_{i}}^{j+1}$, $A_{I_i}^{j+1}$ and $B_{I_i}^{j+1}$, respectively. When $j=n_M+n_K$,
$Y_{I_i}^j=Y_{I_i}$, $A_{I_i}^j=A_{I_i}$ and $B_{I_i}^j=B_{I_i}$.
If we can find $b_j$, $1\leq j\leq n_K+n_M$, such that
for
all $i$, $1 \leq i \leq d$,
\begin{align}\label{sec577}
    \rank(B_{I_i}^j)&=\min\{\row(B_{I_i}^j), \col(B_{I_i}^j)\},
\end{align}
then for
all $i$, $1 \leq i \leq d$, the equality (\ref{sec57}) holds by letting $j=n_K+n_M$ in (\ref{sec577}).


For $1\leq j\leq n_K$, since $B_{I_i}^{j}$ is a sub-matrix of the $n_K\times n_K$
identity matrix $I_{n_K\times n_K}$,
\begin{equation}\notag
    \rank(B_{I_i}^{j})=\row(B_{I_i}^{j}),
\end{equation}
which implies (\ref{sec577}).

Assume that for $j$ equal to some $l\geq n_K$, we have successfully constructed $\{b_i: 1\leq i\leq l\}$ such that for all $i$, $1\leq i\leq d$,
\begin{align}
     \rank(B_{I_i}^l)&=\min\{\row(B_{I_i}^l), \col(B_{I_i}^l)\} \label{eqn6666}
\end{align}
Now in order for (\ref{eqn6666}) to be satisfied with $l+1$ in place of $l$, we need to choose $b_{l+1}$ such that for each wiretap set $I_i$ ($1\leq i\leq d$) containing $x_{l+1}$, if $\row(B_{I_i}^l)<n_K$, then
\begin{equation}\notag
    \rank(B_{I_i}^{l+1}) = \rank(B_{I_i}^{l})+1.
\end{equation}
The existence of $b_{l+1}$ is guaranteed by Lemma \ref{cccc2} provided $q>d$. Then by mathematical induction, $b_j$, $n_K+1\leq j\leq n_K+n_M$, can be chosen as required.

Hence, $b_j$'s are successfully constructed, which completes the
proof.
\end{proof}
For each wiretap set $I_i$,  let $\bar{I}_i=
\mathcal{E}\setminus I_i$. The rate region in Theorem
\ref{ratetuple} can be  rewritten as follows.
\begin{corol}\label{equiva:ratetuple}
    A rate tuple $(R_M,R_K, R_{i:1\leq i \leq d})$ is in $\mathcal{R}'$ if and only if
    \begin{align}
    R_M  & \geq R_i, \ 1\leq i \leq d;\label{ERRRT:2}
    \end{align}
  and there exist $r_i$'s such that
    \begin{align}
     R_K   &= \sum\limits_{i=1}^{h}r_i - R_M; &\label{ERRRT:1}\\
     R_M &\leq \sum\limits_{i=1}^{h}r_i;& \label{ERRRT:4}\\
    0  &\leq r_i \leq C_i,&  \text{$1\leq i\leq h$};\label{ERRRT:5}\\
    0 &\leq R_j \leq  \sum\limits_{e_i\in \bar{I}_j}r_i,& 1\leq j\leq d.\label{ERRRT:6}
    \end{align}
\end{corol}
By comparing the constraints (\ref{ERRRT:2})-(\ref{ERRRT:6}) for $\mathcal{R}'$ and the constraints (\ref{GRRRT:2})-(\ref{GRRRT:5}) for $\mathcal{R}$, we see that they are identical except that (\ref{ERRRT:1}) and  (\ref{GRRRT:1}) are different. Specifically, (\ref{ERRRT:1}) is obtained from  (\ref{GRRRT:1})  by setting the inequality therein to equality. In $\mathcal{R}'$, when $C_i$'s  are fixed,  $r_i, R_M, R_K,$ and  $R_j$'s are all bounded. However, in $\mathcal{R}$, though $r_i$, $R_M$ and $R_j$'s are bounded, $R_K$ can be arbitrarily large. Therefore, $\mathcal{R}'  \subsetneq \mathcal{R}$  in general. However, we will show in Corollary \ref{lastcorl} at the end of the next section that requiring $K$ to be reconstructed  at the receiver by no means impairs the performance of the coding scheme.

\section{The General Rate Region}\label{sec:im4}
In this section, we prove Theorem \ref{General:ratetuple}.
First, we prove the following lemma.

\begin{lemma}\label{lemma:11}
In a cooperative imperfect secrecy system, let  $M$ be the message
and $Y_I$ be the symbols transmitted  in wiretap set $I$. Then
        \begin{equation}\label{star2}
        H(M|Y_I)\leq H(Y_{\bar{I}}|Y_{I}),
        \end{equation}
where $\bar{I} = \mathcal{E}\setminus I$.
\end{lemma}
\begin{proof}Since $\mathcal{E}=I\cup\bar{I}$ and $M$ is a
function of $Y_\mathcal{E}$,
\begin{equation}
H(M|Y_{\mathcal{E}})=0.
\end{equation}
Hence,
    \begin{align}
    H(M|Y_{I})     &= H(M|Y_{I}, Y_{\bar{I}}) + I(M;Y_{\bar{I}}|Y_{I})\notag\\
                   &= I(M;Y_{\bar{I}}|Y_{I})\notag \\
                &\leq H(Y_{\bar{I}}|Y_{I}),\notag
    \end{align}
which completes the proof.
\end{proof}
In this lemma, if we let $I(M;Y_I)=0$, then the inequality (\ref{star2}) reduces to
\begin{equation}
    H(M)\leq H(Y_{\bar{I}}|Y_{I}),
\end{equation}
which is the inequality (\ref{impsec:2}) in  Theorem \ref{thmK}.
\subsection{Converse}
The constraints (\ref{GRRRT:2}) and the left hand side of
(\ref{GRRRT:5}) can be proved by the the same method in  Section
\ref{sec1:converse}.  Let us focus on the remaining constraints.

Since
        $Y_{(e_i:1\leq i\leq h)}$ is a function of $(M,K)$,
        \begin{equation}\label{geqineq:1}
        H(Y_{(e_i:1\leq i\leq h)})\leq  H(M,K)=H(M)+H(K).\notag
        \end{equation}
        Hence,
        \begin{equation}\notag
        H(K) \geq H(Y_{(e_i:1\leq i\leq h)}) - H(M),
        \end{equation}
        which is equivalent to
        \begin{equation}\label{ineq:l1}
        R_K \geq H(Y_{(e_i:1\leq i\leq h)}) - R_M.
        \end{equation}
Since
         $M$ can be recovered from $Y_{(e_i:1\leq i\leq h)}$,
        \begin{equation}\label{geqineq:2}
        H(Y_{(e_i:1\leq i\leq h)})\geq  H(M),\notag
        \end{equation}
        which is equivalent to
        \begin{equation}\label{ineq:l2}
        R_M \leq H(Y_{(e_i:1\leq i\leq h)}).
        \end{equation}
For any wiretap set $I_i$,  $1\leq i\leq d$,
      $$Y_{(e_i:1\leq i\leq h)} = Y_{(I_i, \bar{I}_i)}.$$
By the constraint (\ref{eqn:equivacation}) and Lemma \ref{lemma:11},
for all $1\leq j\leq d$,
\begin{equation}\label{ineq:l3}
R_j \leq H(M|Y_{I_j}) \leq H(Y_{\bar{I}_j}|Y_{I_j}).
\end{equation}
        For $1\leq i \leq h$, let $$r_i = H(Y_{e_i}|Y_{(e_1, e_2, ..., e_{i-1})}).$$
        Then
    $$r_i\leq H(Y_{e_i}|Y_{(e_l: e_l\in I_j, l< i)}).$$
        Furthermore, (\ref{ineq:l1})  implies
    \begin{align}
     R_K &\geq  H(Y_{(e_i:1\leq i\leq h)}) - R_M  \notag\\
    &= \sum\limits_{i=1}^{h}H(Y_{e_i}|Y_{(e_1, e_2, ..., e_{i-1})}) -R_M\notag\\
    &=\sum\limits_{i=1}^{h} r_i -R_M,\notag
    \end{align}
     and  (\ref{ineq:l2}) implies
    \begin{align}
     R_M &\leq H(Y_{(e_i:1\leq i\leq h)})\notag\\
        &= \sum\limits_{i=1}^{h}H(Y_{e_i}|Y_{(e_1, e_2, ..., e_{i-1})})\notag\\
        &=\sum\limits_{i=1}^{h}r_i.\notag
        \end{align}
 Also,
 \begin{align}
 0&\leq r_i \leq H(Y_{e_i}) \leq C_i.\notag
 \end{align}
Finally, (\ref{ineq:l3})\text{ implies }
\begin{align}
R_j &\leq H(Y_{\bar{I}_j}|Y_{I_j}) \notag \\
     & = \sum\limits_{e_i\in \bar{I}_j} H(Y_{e_i}|Y_{(e_l: l<i)}, Y_{I_j})\notag\\
     &\leq \sum\limits_{e_i\in \bar{I}_j} H(Y_{e_i}|Y_{(e_l: l<i)})\notag\\
     &= \sum\limits_{e_i\in \bar{I}_j} r_i.\notag
\end{align}
 Hence,  we prove all the constraints in (\ref{GRRRT:2})-(\ref{GRRRT:5}).

\subsection{Achievability}
In the above converse, the only constraint on $R_K$ is
\begin{equation}
     R_K   \geq \sum\limits_{i=1}^{h}r_i - R_M.\notag
\end{equation}
 Let $\hat{R}_K=\sum\limits_{i=1}^{h}r_i - R_M$ and fix $R_M$ and $R_{i},1\leq i\leq d$.
From the discussion in Section \ref{sec:ach}, the rate tuple
$(R_M, \hat{R}_K, R_{i:1\leq i\leq d})$ can be attained. Then $(R_M,
R_K, R_{i:1\leq i\leq d})$ can be attained  by discarding
$R_K-\hat{R}_K$ bits of the key before constructing a code for $(R_M,
\hat{R}_K, R_{i:1\leq i\leq d})$. Hence we have the following corollary, which shows  that requiring $K$ to be reconstructed by at the receiver by no means impairs the performance of the coding scheme.

\begin{corol}\label{lastcorl}
Fix $R_M$ and $R_{i:1\leq i\leq d}$ in a rate tuple ($R_M, R_K,
R_{i:1\leq i\leq d}$), if $R_K$ is minimized then $K$ can be
recovered by the receiver.
\end{corol}

\section{Stochastic Encoder}\label{sec:stochastic-encoder}
\begin{figure}
\begin{center}
\includegraphics[height= 200pt, width = 500pt]{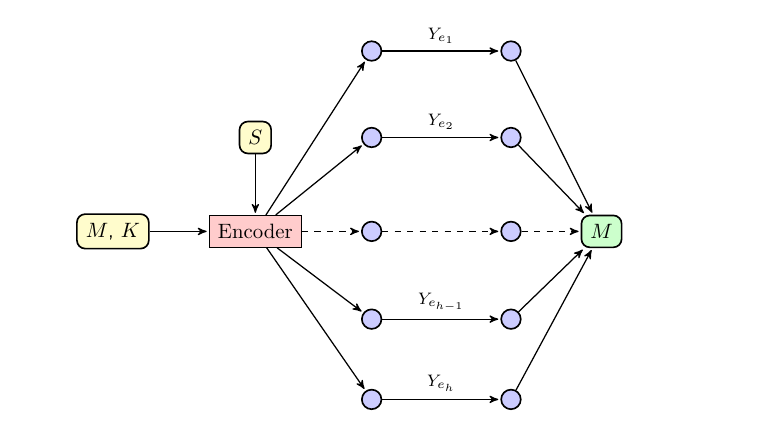}
\caption{The Stochastic Encoder.}\label{fig-im2}
\end{center}
\end{figure}
We have already established the rate region when the encoding at the sender is deterministic, i.e., the information symbols on all the channels are a function of $M$ and $K$. In this section,   in stead of the deterministic encoder, a stochastic encoder is employed at the sender, where the  information symbols on all the channels are no longer a function of $M$ and $K$. Hence, the size of $K$ is not of our concern as extra randomness is injected by the stochastic encoder. We continue to assume that no randomness is introduced inside the network. We show that under this more general model,
 the characterization of  the message rates  remains the same as that in Theorem \ref{General:ratetuple}.

The communication model is depicted in Fig. \ref{fig-im2}. The problem statement is almost the same as that in Section \ref{sec:im1}, with the only exception that the symbols on the channels is determined by a stochastic matrix. In principle, a stochastic encoder can be equivalently transformed into a deterministic encoder by introducing an auxiliary random variable which is independent of the encoder input (cf. p. 141, Yeung \cite{yeung2008information}).  Denote the auxiliary random variable in the block code  by $S^n$, which may depend on $M^n$ and $K^n$.

Now, we  summarize the  conditions that  hold when the encoder may be stochastic.
At the sender,
\begin{equation}
I(M^n; K^n) = 0.\label{stoce-2}
\end{equation}
The information symbols on the channels satisfy that
\begin{equation}
H( Y_{\mathcal{E}}^n | M^n, K^n, S^n) = 0.\label{stoce-3}
\end{equation}
Since the message $M^n$ can be decoded at the receiver,
\begin{equation}\label{decoding-m}
H(M^n|Y_{\mathcal{E}}^n) = 0.
\end{equation}
As required, for each wiretap set $I_j$,
\begin{equation}
\frac{H(M^n|Y_{I_j}^n)}{n}\geq R_j, 1 \leq j \leq d. \label{stoce-1}
\end{equation}
Next, we show that the message rate cannot be increased by using a stochastic encoder, i.e., conditions (\ref{GRRRT:2}) and (\ref{GRRRT:3}) -- (\ref{GRRRT:5}) continue to hold.

It is easy to verify condition (\ref{GRRRT:2}).  Close examination of the proof of  Lemma \ref{lemma:11} reveals that the lemma remains valid in light of \eqref{stoce-2}--\eqref{decoding-m} (in fact the proof depends only on \eqref{decoding-m}).

For $1\leq i \leq h$, let $$r_i = H(Y_{e_i}^n|Y_{(e_1, e_2, ..., e_{i-1})}^n)/n.$$
Then
    $$0\leq r_i\leq H(Y_{e_i}^n|Y_{(e_l: e_l\in I_j, l< i)}^n)/n\leq H(Y_{e_i}^n)/n\leq C_i,$$
which is condition (\ref{GRRRT:4}).
By \eqref{stoce-1} and  Lemma \ref{lemma:11},
\begin{align}
R_j  &\leq  H(M^n|Y^n_{I_j})/n \notag\\
     &\leq H(Y_{\bar{I}_j}^n|Y_{I_j}^n)/n \notag \\
     & = \sum\limits_{e_i\in \bar{I}_j} H(Y_{e_i}^n|Y_{(e_l: l<i)}^n, Y_{I_j}^n)/n \notag\\
     &\leq \sum\limits_{e_i\in \bar{I}_j} H(Y_{e_i}^n|Y_{(e_l: l<i)}^n)/n\notag\\
     &= \sum\limits_{e_i\in \bar{I}_j} r_i,\notag
\end{align}
which is condition (\ref{GRRRT:5}).
By  (\ref{decoding-m}),
$$R_M=H(M^n)/n \leq H(Y_{\mathcal{E}}^n)/n=\sum\limits_{i=1}^{h}r_i,$$
which is condition (\ref{GRRRT:3}).

Hence, the message rate cannot be increased by introducing stochastic encoding at the sender.

\section{Conclusion}
In this paper, we have obtained a tight rate region for the cooperative imperfect secrecy model in terms of  a linear program, of
which the key idea is from the imperfect secrecy theorem.  Although
the rate region  is still open for the general case, our work has paved
the way for further investigation into this problem.


\end{document}